\documentclass[]{article}
\usepackage[a4paper, margin=1in]{geometry}
\usepackage[utf8]{inputenc}
\usepackage[T1]{fontenc}
\usepackage{graphicx}
\usepackage{bm}
\usepackage{siunitx}
\usepackage{authblk}
\usepackage{xcolor}
\usepackage{cite}
\usepackage{lineno}

\begin{document}
\title{Coherent phase transfer for real-world twin-field quantum key distribution
}
\author[1,*]{Cecilia Clivati}
\author[1,2]{Alice Meda}
\author[1]{Simone Donadello}
\author[1,3]{Salvatore Virzì}
\author[1]{Marco Genovese}
\author[1]{Filippo Levi}
\author[1]{Alberto Mura}
\author[4,5]{Mirko Pittaluga}
\author[4]{Zhiliang L. Yuan}
\author[4]{Andrew J. Shields}
\author[6]{Marco Lucamarini}
\author[1,2]{Ivo Pietro Degiovanni}
\author[1]{Davide Calonico}
\affil[1]{INRIM, strada delle cacce 91, 10134 Torino, Italy}
\affil[2]{INFN, sezione di Torino, via P. Giuria 1, 10125 Torino, Italy}
\affil[3]{Physics Department, University of Torino, via P. Giuria 1, 10125 Torino, Italy}
\affil[4]{Toshiba Europe Ltd, 208 Science Park, Milton Rd, Cambridge CB40GZ, U.K.}
\affil[5]{School of Electronic and Electrical Engineering, University of Leeds, Leeds LS2 9JT, U.K.}
\affil[6]{Department of Physics and York Centre for Quantum Technologies, University of York, YO10 5DD York, U.K.}
\affil[*]{c.clivati@inrim.it}
\maketitle
\begin{abstract}
Quantum mechanics allows the distribution of intrinsically secure encryption keys by optical means. Twin-field quantum key distribution is the most promising technique for its implementation on long-distance fibers, but requires stabilizing the  optical length of the communication channels between parties. 
In proof-of-principle experiments based on spooled fibers, this was achieved by interleaving the quantum communication with periodical adjustment frames. In this approach, longer duty cycles for the key streaming come at the cost of a looser control of channel length, and a successful  key-transfer using this technique in a real world remains a significant challenge. 
Using interferometry techniques derived from frequency metrology, we developed a solution for the simultaneous key streaming and channel length control,  and demonstrate it  on a \SI{206}{km} field-deployed fiber with \SI{65}{dB} loss. 
Our technique reduces the quantum-bit-error-rate contributed by channel length variations to <1\%,  representing an effective solution for real-world quantum communications.
\end{abstract}





\begin{figure}
    \centering
    \includegraphics[width=0.5\textwidth]{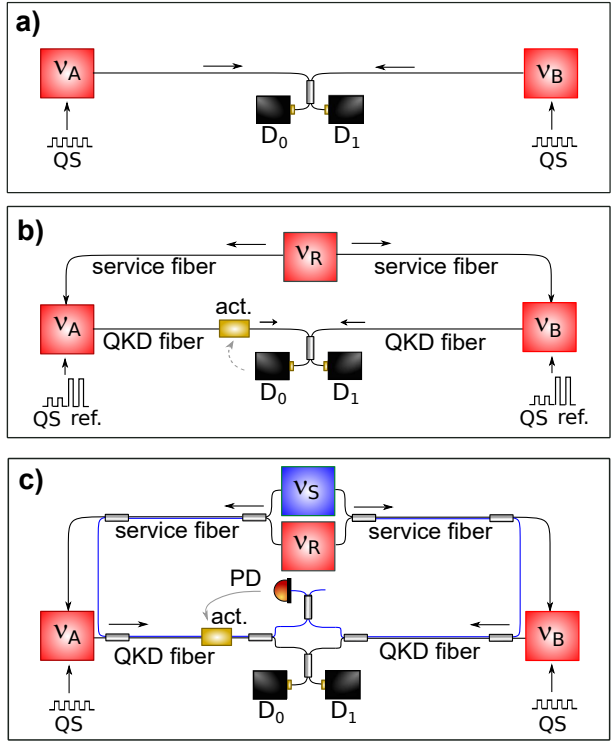}
    \caption{
   \textbf{Principle schemes of TF-QKD.} a) In ideal TF-QKD, Alice and Bob encode quantum states (QS) on local lasers, attenuated to the single-photon level and with equal frequencies $\nu_\text{A}=\nu_\text{B}$. The resulting signals are sent to Charlie, where they interfere on single-photon detectors (D\textsubscript{0} and D\textsubscript{1}). b) In practical implementations, a reference laser with frequency $\nu_\text{R}$ is sent to Alice and Bob through a service fiber, to  phase-lock the QKD lasers and ensure $\nu_\text{A}=\nu_\text{B}=\nu_\text{R}$. After information encoding, QKD lasers are sent to Charlie through the QKD fibers, whose optical-length-changes are detected by interleaving the key streaming with classical transmission. Optical length fluctuations are counteracted by adjusting the phase of the lasers through an actuator (act.). c) In our approach, an additional sensing laser, with frequency $\nu_\text{S}$ travels the service fiber with the reference laser, and the QKD fibers together with the QKD lasers. It can be spectrally separated  because $\nu_\text{S}$ falls in a different channel of the wavelength-division-multiplexing grid. While QKD lasers interfere on D\textsubscript{0} and D\textsubscript{1}, the classical signals at $\nu_\text{S}$ are phase-compared on a photodiode (PD) to detect the noise of both the service and QKD fibers. This allows tight control of the fiber noise and simultaneous key streaming.
    \label{fig:setup_gen}}
\end{figure}

Quantum key distribution (QKD) enables to share secret cryptographic keys between distant parties, whose intrinsic security is guaranteed by the laws of quantum mechanics \cite{bennet, scarani, lo}. 
Besides pioneering experiments involving satellite transmission \cite{pan1,pan2}, the challenge is now to integrate this technology on the long-distance fiber networks already used for telecommunications  \cite{shimizu,zavatta,wenger,mao,tanaka,choi, dixon,peev,sasaki,dynes}. The maximum secure key rate for QKD decreases exponentially with the channel losses. Although the reach could be extended using quantum repeaters, the related research is still at a rudimentary level and these devices are far from  operational  \cite{fxu,briegel, sangouard}. Nowadays, intercity distances could only be covered using trusted nodes \cite{peev}, whose security represents however a significant  technical issue. A fundamental resource for next-generation long-distance secure communications is represented by the recently proposed twin-field QKD (TF-QKD) protocol \cite{lucamarini}, because of its weaker dependence on  channel loss. \\
In TF-QKD, the information is encoded as  discrete phase states on dim laser pulses generated at distant Alice and Bob terminals and sent through optical fiber to a central node, Charlie, where they interfere. 
This idea, sketched in Fig. \ref{fig:setup_gen}a, was proved secure against general attacks 
\cite{ma,wangx,lin,curty,cui} also in the finite-size scenario \cite{yu, yin,lorenzo} and with the aid of two-way communication \cite{xu}, but it is based on the critical assumptions that the optical pulses are phase-coherent in Alice and Bob and preserve coherence throughout the path to Charlie. 
While the first requirement can be fulfilled by phase-locking the two QKD lasers in Alice and Bob to a common reference laser transmitted through a service channel, the uncorrelated  fluctuations of the length and refractive index of the connecting paths (i.e. the optical length) introduce phase noise to the system and reduce the visibility of the interference measurement. In proof-of-principle experiments based on spooled fibers \cite{wang, minder,zhong, chen,fang}, this effect was mitigated by interleaving the  QKD frames with classical transmission frames that  were used to  periodically realign the phases of interfering pulses  \cite{wang,minder} (see  Fig. \ref{fig:setup}b). However, this approach becomes less effective as the length  of connecting paths exceeds few hundreds of kilometers  \cite{wang, chen,fang}
 and there is no experimental evidence that it could work in  deployed fibers,  where the attenuation and phase fluctuations are considerably higher  \cite{clivatiOPTICA2018}. \\
 We propose a solution derived from frequency metrology. In this context,  the transmission of  coherent laser radiation over thousand-kilometer-long fibers is employed to compare distant atomic clocks at the highest accuracy \cite{clivatiVLBI,grotti, lisdat,delva,guena,katori,bacon}. This is made possible by  the use of  ultrastable lasers and the active cancellation of the noise introduced by connecting fibers \cite{williams}, and the same approach can be exploited in TF-QKD.\\
We realised an apparatus suitable for TF-QKD where the phase fluctuations of both the lasers and connecting fibers are actively cancelled. This is achieved by transmitting  an additional laser in the same fiber as the QKD lasers in a wavelength-multiplexed approach. In Charlie, this is used to sense and stabilize the channels' optical length variations (see Fig. \ref{fig:setup_gen}c). In a QKD experiment, this allows simultaneous key streaming and channels stabilization, ensuring  longer duty-cycles and  a tighter control of the optical phase on long-haul deployed fibers,  where interleaved approaches would fail. We implement our solution  on a real-world network where the distance between Alice and Bob is \SI{206}{km} and the net losses are as high as \SI{65}{dB}, demonstrating a significant progress over existing quantum communication field trials \cite{shimizu,zavatta,wenger,mao,tanaka,choi,dixon,peev,sasaki,dynes}, all limited to \SI{<100}{km} distance and \SI{<25}{dB} channel loss. \\

\section*{The experiment}
\subsection*{Experimental setup}
\begin{figure}
    \centering
    \includegraphics[width=\textwidth]{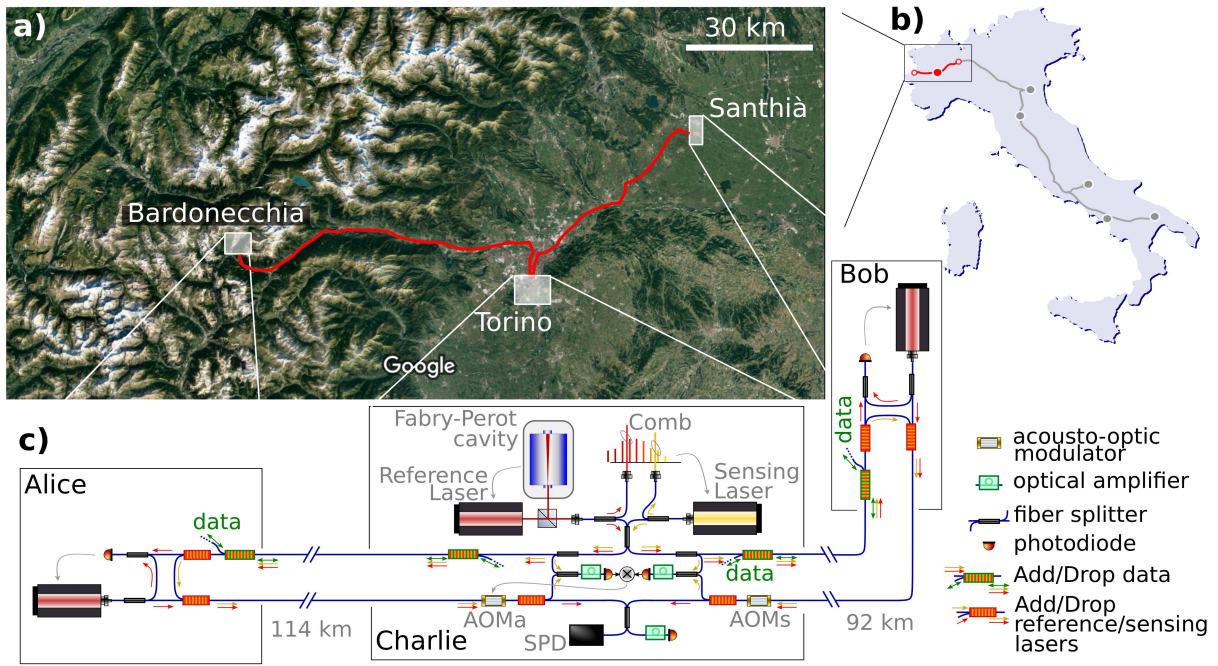}
    \caption{\textbf{Map and experimental set-up. } a) Charlie was located at INRIM (Torino), while Alice and Bob were located in shelters of the telecom network in Bardonecchia and Santhià (Imagery \copyright 2020 Landsat/Copernicus, Imagery \copyright 2020 TerraMetrics, Map data ©2020). b) A sketch of the Italian Quantum Backbone, with the spans used in this experiment coloured in red and the red-filled (empty) circles representing Charlie (Alice and Bob). c) The experimental layout. The reference laser in Charlie is stabilized to a high-finesse cavity and the sensing laser is phase-locked to it using an optical comb. Wavelength division multiplexers combine/separate them and add/drop bidirectional data signals. In Alice and Bob, we detect the beat between the incoming reference laser and a local laser and phase-lock the two. The QKD laser light is recombined with the sensing laser and sent to Charlie on a dedicated fiber. The acousto-optic modulator AOMa adjusts the optical phase to correct for the  noise introduced by the fiber optical length variations, while AOMs is a fixed frequency shifter. The fiber noise is detected by interfering the local sensing laser with return light from each arm. The two beatnotes are detected on separate photodiodes and phase-compared in the RF domain. The QKD lasers interfere on a photodiode when performing experiments in the classical regime, and on a single photon detector (SPD) in the photon-counting regime, also used to detect background photons.}
    \label{fig:setup}
\end{figure}

The map and detailed scheme of our experiment are shown in Fig. \ref{fig:setup}. We use a pair of ultrastable lasers
with a linewidth of  $\sim$\SI{1}{Hz} and frequency $\nu_\text{R}=$\SI{194.4}{THz} (\SI{1542.14}{nm}) and $\nu_\text{S}=$\SI{194.25}{THz} (\SI{1543.33}{nm}), which are standard frequencies of the dense wavelength-division multiplexed (DWDM) grid. 
The former (hereafter, reference laser) is used as a reference for locking the QKD lasers in Alice and Bob terminals, and  is frequency stabilized to a high-finesse ultrastable Fabry-Perot cavity \cite{clivatiuffc}. The latter (hereafter, sensing laser) is used to detect the fiber noise and allows its cancellation. In our experiment, we  offset-locked it to the reference laser using an optical frequency comb  \cite{telle}, although alternative techniques are possible (see Methods). 
They are combined using a commercial \SI{100}{GHz}-wide DWDM filter and sent  to Alice and Bob through separate service fibers. \\
At the remote terminals, the reference laser is extracted and used to phase-lock the local QKD laser. 
This is recombined with the sensing laser  and  sent back to Charlie on the QKD fiber. This setup implements what  is needed for transmitting quantum information,  nonetheless we do not realise a fully-operative QKD transmission  since this is a technical issue already demonstrated elsewhere \cite{wang,minder,zhong,chen}. Instead, our experiment focuses on improving   the system coherence, which is the essential prerequisite for any TF-QKD protocol. \\
In Charlie, we interfered the  QKD lasers in classical and photon-counting regimes, the latter after  attenuating the QKD lasers to the single-photon level.  
 As the sensing laser travels the path from Charlie to Alice and Bob together with the reference laser on the service fiber, and the backward path together with the QKD laser on the QKD fiber, its accumulated phase contains information on the optical length changes of connecting paths, that can be used to stabilize them. The incoming beams at the two wavelengths are  routed to separate detectors: photons from the QKD lasers interfere on a photodiode (when doing experiments in the classical regime) or a single-photon  detector (SPD, in the photon counting regime),  while the sensing laser beams are indirectly phase-compared using a pair of photodiodes and a radio-frequency mixer. From it, the relative changes in the optical paths are extracted and stabilised by a phase-locked loop. This is achieved by adjusting the phase of the sensing laser as it travels the acusto-optic modulator AOMa, on Alice's branch. Because the QKD laser also travels through the actuator, its optical phase is stabilised as well  (see Methods). \\
We implemented this scheme over long-haul fiber backbones connecting INRIM, in the city of Torino (Italy), where the Charlie terminal was located, to network nodes separated by \SI{114}{km} and \SI{92}{km} of optical fiber with \SI{35}{dB} and \SI{30}{dB} losses (Alice and Bob terminals respectively). The overall length of the fiber connecting Alice and Bob was thus \SI{206}{km} with an  attenuation as high as \SI{65}{dB}. The average loss coefficient of \SI{0.3}{dB/km} is higher than the specified level for standard optical fibers (\SI{0.2}{dB/km}) and includes discrete losses of the connectors and DWDM  equipment, which play a significant role in deployed networks.
These fibers are part of the Italian Quantum Backbone  and carry other services, among which is the dissemination of  atomic clock signals to research facilities of the Country  \cite{clivatiVLBI,grotti}. While conventional fiber communication protocols are based on
data exchange over fiber pairs, in which each fiber allows light propagation in a single direction, we implemented a bidirectional  transmission  on a single fiber, using different DWDM channels for  opposite directions. 
Using this approach, the second fiber of the pair was dedicated to the sensing and QKD lasers only (see Methods).   The service fibers carry, in addition to the reference and sensing lasers, also standard data traffic and  time/frequency information, which can be conveniently used to implement ultra-precise modulation of the quantum signals and transmit the classical information typical of any QKD protocol, including
TF-QKD.  Here, we  used a White Rabbit precise time protocol \cite{serrano,dierikx} to distribute clock information for the optical phase-lock of the QKD laser in Bob.

\subsection*{Results}
\begin{figure}
    \centering
     \includegraphics[width=0.9\textwidth]{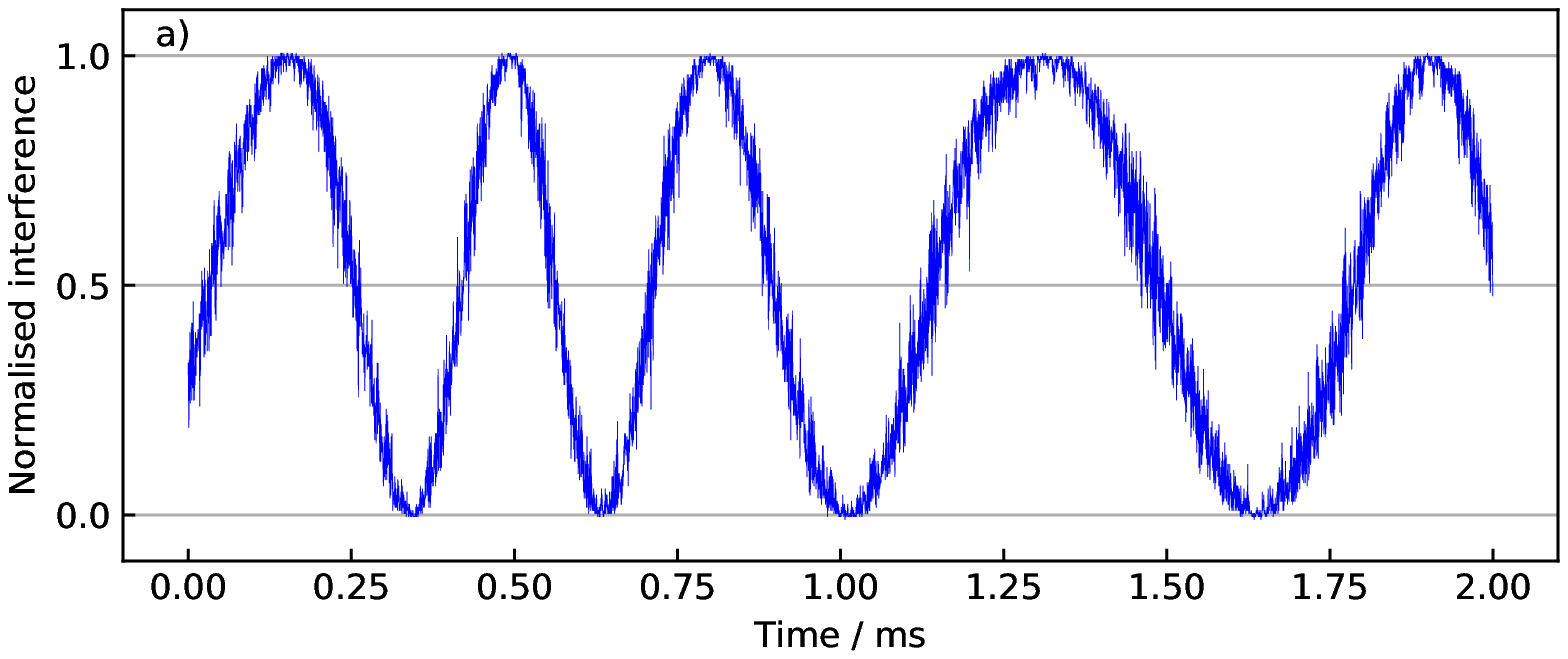}\\
  \includegraphics[width=0.9\textwidth]{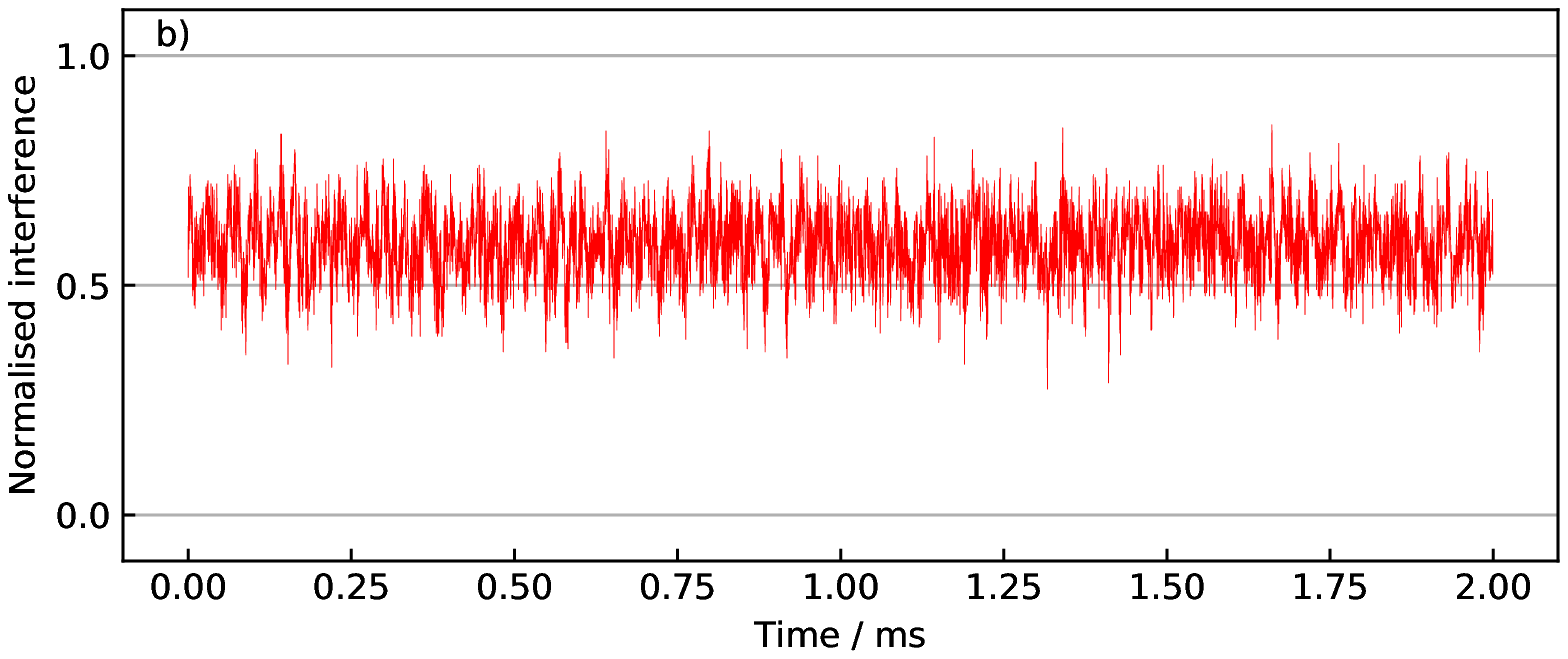}\\
    \includegraphics[width=0.9\textwidth]{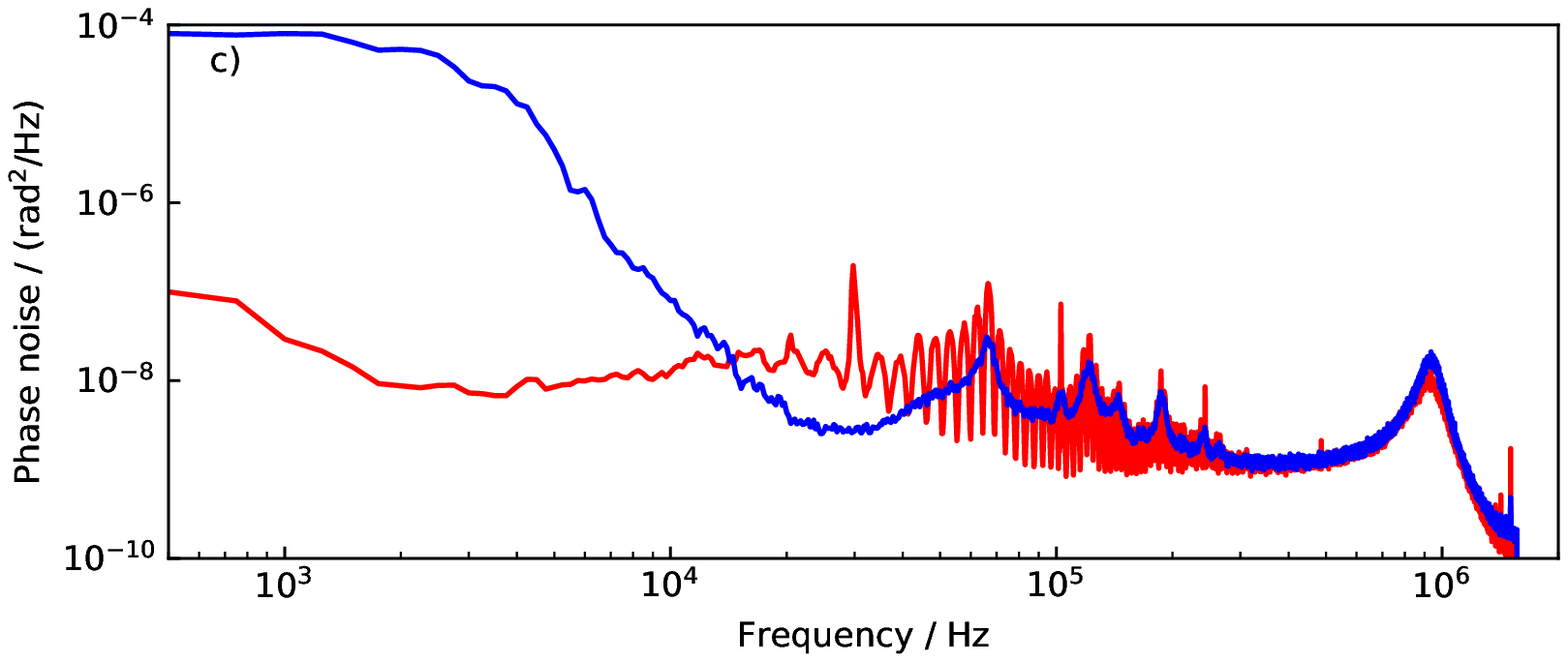}\\
    \caption{\textbf{QKD lasers interference with unstabilized and stabilized fibers}. We record the interference between the QKD lasers in Charlie on a fast photodiode (the traces are normalised between 0 and 1. a) In an unstabilized condition the instantaneous phase drifts by 30 rad/ms and is folded back when it exceeds the $[0 :\pi]$ interval. b) When the fiber is stabilized, the phase remains stable. In this measurement, the interferometer was stabilized far from the folding point, i.e. in a condition where phase fluctuations were directly mapped into intensity fluctuations, to investigate the residual noise processes. c) The power spectral density of the phase. A significant reduction in the noise is observed in a stabilized condition (red) with respect to an unstabilized condition (blue). The apparent plateau observed at low Fourier frequencies in an unstabilized condition is originated by the folding of the interferometer response. At high Fourier frequency, similar noise is observed in the two traces, mainly due to the QKD lasers  and the self-delayed interference of the reference and sensing lasers.}
    \label{fig:vis}
\end{figure}

\begin{figure}
    \centering
    \includegraphics[width=0.8\textwidth]{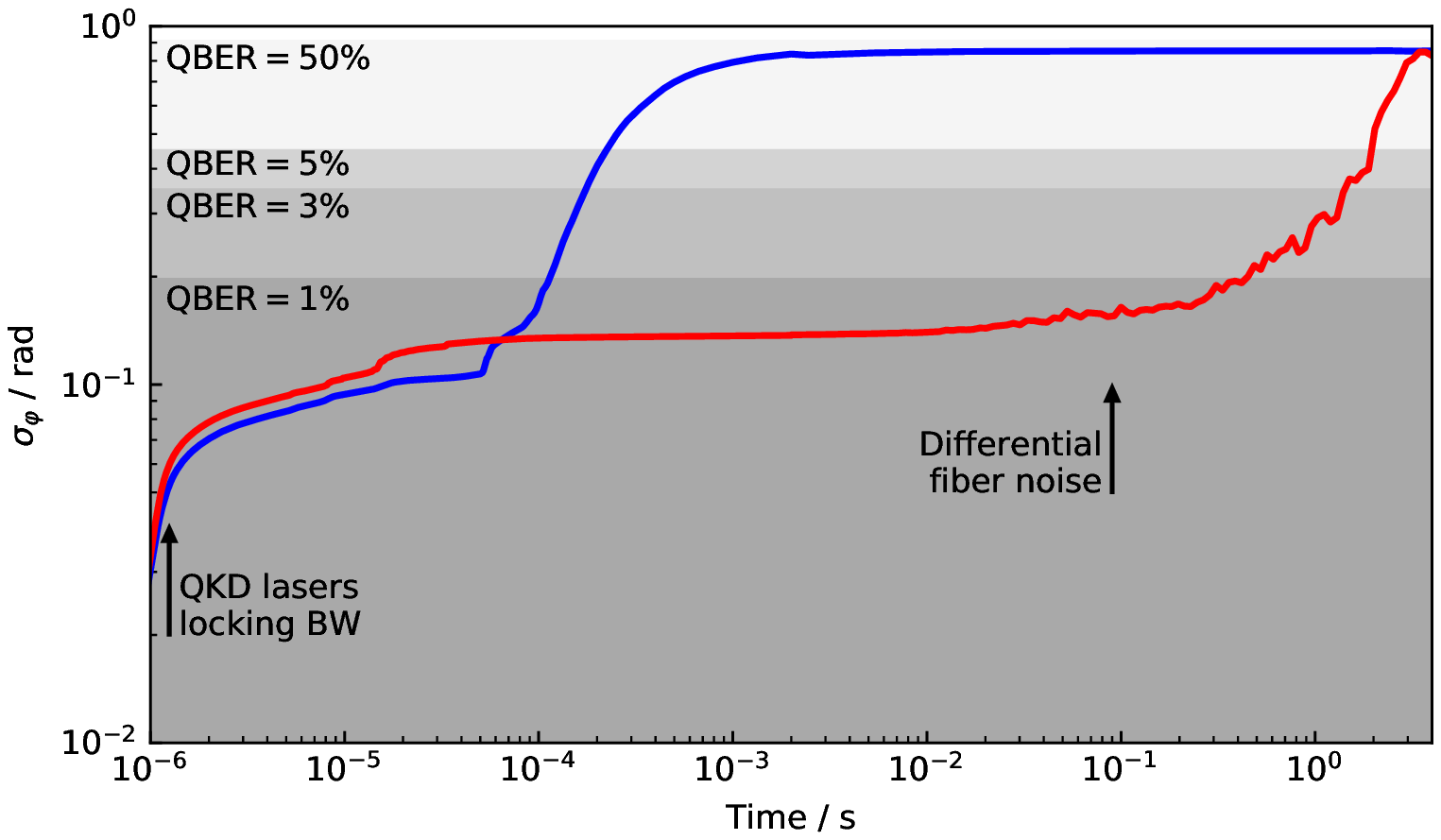}
    \caption{\textbf{Phase fluctuations over time.} The deviation of the phase $\sigma_\varphi$ between the two QKD lasers interfering in Charlie at different timescales, in an unstabilized (blue) and stabilized (red) condition. For this calculation, we acquired the interference pattern over \SI{4}{s} and subdivided it in shorter time frames, calculating  $\sigma_\varphi$ for each frame. The shadowed areas indicate upper thresholds for relevant values of the QBER, quantified as $\sigma^2_\varphi/4$ in the low-noise ($\varphi\approx 0$) approximation. The arrows indicate timescales where the QKD lasers noise and the uncompensated fiber noise (uncorrelated fluctuations at the two wavelengths) become relevant. The phase, and corresponding QBER, were retrieved from the interference pattern according to the procedures described in the Methods.}
    \label{fig:sigma}
\end{figure}

\begin{figure}
    \centering
    \includegraphics[width=\textwidth]{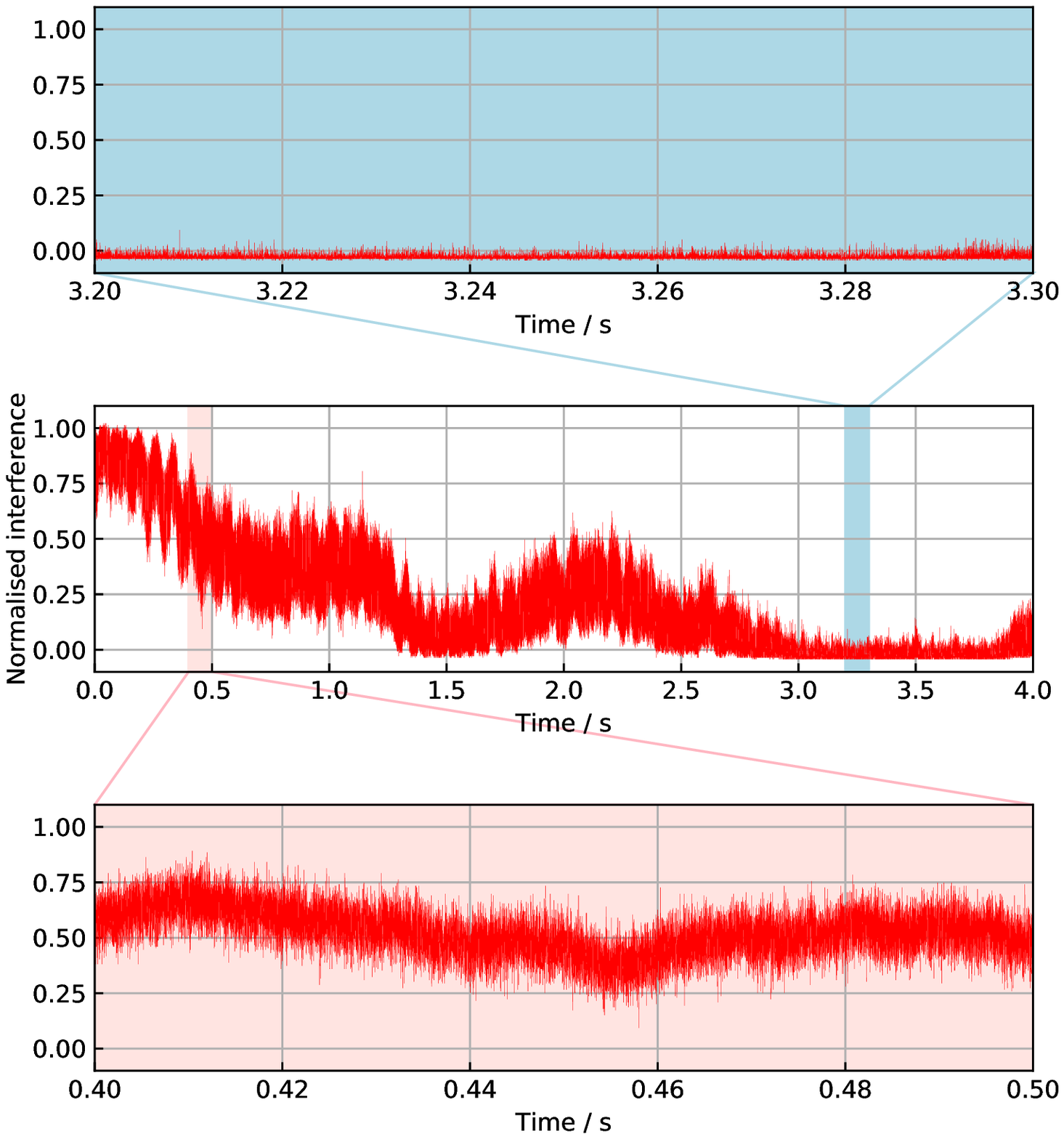}
    \caption{\textbf{QKD lasers interference on the long term.} In the central panel, the normalised interference pattern in the stabilized-fiber condition is shown in a \SI{4}{s}-long time window. The blue shadowed area in the uppermost panel indicates a \SI{100}{ms}-long fraction of the time interval in which the interferometer operated in the maximum visibility condition, i.e. the one exploited for the exchange of the bits of the cryptographic key.  The red shadowed area in the lowermost panel indicates a \SI{100}{ms}-long time interval in which the interferometer operated around the 0.5 visibility. Configurations far from the deterministic behavior of the interfermeter (i.e. far from the maximum and minimum of the visibility) are the ones where the relation between the  phase fluctuations and the QBER is linear, and could be used to realign the  phase on the long term and mitigate the residual uncontrolled optical path length variations. }
    \label{fig:vislong}
\end{figure}

Fig. \ref{fig:vis} shows the interference between the QKD lasers  measured on a photodiode in Charlie  in a \SI{2}{ms} time frame, without (a, blue) and with (b, red) active stabilization of the fiber paths. 
In an unstabilized condition, several  phase cycles are accumulated in the considered interval, with an instantaneous drift of up to \SI{30}{rad/ms}.  
When the path is stabilized, on the contrary, the  phase remains stable over the whole acquisition frame. In this measurement, the phase was stabilized on purpose at a point where its fluctuations were directly mapped into intensity fluctuations, which enabled us to compute the corresponding phase  noise power spectral density.    
This is shown in Fig. \ref{fig:vis}c: in an unstabilized condition (blue) the phase noise rapidly diverges at low Fourier frequencies, while, when stabilization is activated (red), the noise is suppressed up to a bandwidth of tens of kilohertz. 
In both traces, the noise floor is set by  the QKD lasers at and within the locking bandwidth of \SI{0.9}{MHz} and the self-delayed interference of the reference and  sensing lasers. These contributions are common to the two traces. The  latter becomes proportionally higher as the length unbalance between the arms of the interferometer increases and less stable lasers are used. The use of ultrastable lasers was crucial in our setup, where the  unbalance was \SI{22}{km}, i.e. \SI{44}{km} of differential path considering both the service and QKD fibers. 
As the relevant noise processes  extend up to $\sim$\SI{1}{MHz} Fourier frequency, we note that an acquisition system with a minimum measurement bandwidth of \SI{2}{MHz} is required to fully capture them. Devices with slower frequency response would act as  low-pass filters,  leading to underestimation of the corresponding phase changes.   \\
The QBER associated to  phase-decoherence  can be calculated from the standard deviation of the phase $\sigma_\varphi$ as a function of the frame duration  (see Methods).
Figure \ref{fig:sigma} shows the results in a stabilized (red) and unstabilized condition (blue).
In both cases, the noise processes responsible for phase fluctuations  extinguish at timescales shorter than the inverse of the locking bandwidth of the QKD lasers (indicated by the arrow), making $\sigma_\varphi$ negligible. 
  However, in  an unstabilized condition, the system exceeds the  1\% QBER threshold in about \SI{100}{\micro s}. 
  At timescales longer than a few milliseconds, apparently, the phase fluctuations do not increase further. This is an artifact caused by the limited range of the interferometer response, which wraps the phase into the $[0;\pi]$ interval. In practice, the  phase wanders by tens of radians in few milliseconds.   
  When stabilization is activated, instead, the system settles in a condition where   $\sigma_\varphi$ = \SI{0.13}{rad}, which corresponds to  a QBER of 0.5\%,  for about \SI{100}{ms}. This value is  determined by the residual contribution of the reference, sensing and QKD lasers noise. For a frame duration longer than \SI{100}{ms},  $\sigma_\varphi$ increases due to a  non perfect cancellation of the fiber optical length variations. This depends on the fact that such variations are detected through the accumulated phase of the sensing laser, while the reference and QKD lasers accumulate a slightly different phase because of the wavelength  difference and uncommon optical paths. This effect is largely predictable and  could be  reduced with dedicated electronics and an optimised design of the experimental setup, thus allowing further extension of the coherence-time.\\
   The observed residual phase noise and QBER  represent conservative estimates, as our measurements were performed on a testbed with as much as \SI{22}{km} of unbalance between the two arms and with standard telecom diode lasers at the Alice and Bob terminals.
   Further improvement could be gained using less noisy telecom lasers \cite{narrowlaser,maleki} and faster  control techniques. However, already in the present condition, the system can be operated at a QBER < 3\% for timescales of the order of \SI{1}{s}. Figure \ref{fig:vislong} shows the interference pattern on a \SI{4}{s} timescale, and a zoom of a \SI{1}{s}-long period where the system could be operated at the maximum  visibility in a QKD experiment. We also show a zoom of a \SI{100}{ms}-long region where the interferometer operates far from the deterministic condition.  With such a stability, in a QKD experiment, it becomes possible to gather enough photon statistic for  realigning the phase  on the basis of the  QBER, thus virtually ensuring even longer duty cycles.  \\
  The same measurements were repeated by attenuating the QKD lasers at the remote terminals by $\sim$\SI{80}{dB}, so that only few thousands of photons/s reach the detector in Charlie, under similar operating conditions  as in recent TF-QKD experiments \cite{wang,minder, zhong,chen}. For this measurement, we replaced the photodiode with an SPD and recorded the number of counts as a function of time.  We were able to reproduce the same visibility as with the classical beams, showing substantial agreement between the two approaches. \\
In view of the implementation of this technique in a QKD experiment, an aspect of concern is the control of the background photons which couple to the QKD fibers from the surrounding environment, or are originated in the QKD fibers themselves due to nonlinear effects. 
Only photons at the QKD lasers wavelength  are relevant to the count, as those in other bands can be filtered out. To counteract the drop in performances of standard DWDM filters outside the C-band, we combined them  with broader  filters featuring  \SI{50}{dB} attenuation throughout the visible and near infrared spectrum (see Methods). 
This ensures efficient separation of the sensing and QKD lasers photons, and provides adequate immunity to background photons from external sources even when the network occupancy and its spectral distribution are unknown. \\
Background photons in the same  band as the quantum signal are mainly produced by  the Raman scattering of the sensing laser in the QKD fiber. This problem is well known in the context of real-world QKD and forces to use either dedicated fibers or QKD transmission in the O-band at \SI{1310}{nm}, where the scattering from  channels in the C-band is negligible \cite{mao,gobby}. In our experiment, we could  minimise its impact by ensuring that the sensing laser power coupled to the QKD fiber at the Alice and Bob terminals was $<$\SI{1}{\micro W}. 
Another effect contributing photons in the same band as QKD lasers is the Rayleigh scattering of the reference laser happening in the service fiber. Rayleigh-scattered photons propagating backward into the service fiber may fall into the QKD fiber due to evanescent coupling. 
In our case,  the maximum allowed power for the reference laser was \SI{20}{\micro W}, which was still enough for ensuring a stable phase lock of the slave  lasers at Alice and Bob terminals. Because this effect is stronger in the first $\sim$ \SI{20}{km} from Charlie,  in case of longer distances or more lossy fibers,  Rayleigh scattering could be kept negligible by maintaining the same level of  launched power, and exploiting  optical amplification closer to remote Alice and Bob terminals.\\
We measured the background photon rate in our setup exploiting a low noise InGaAs/InP avalanche photodiode  with quantum efficiency of 10\% and dead time of \SI{25}{\micro s}.
In the working conditions the observed rate of background photons was  $(5.09 \pm 0.01)\, \text{s}^{-1}$, evaluated over 24 h of measurement, primarily attributed to Raman scattering of the sensing laser.  When all the laser sources involved in our experiment were switched off, the measured level of background photons was $(4.76 \pm 0.04)\, \text{s}^{-1}$, slightly above the intrinsic dark count rate of our SPD, i.e. $(4.52 \pm 0.03)\, \text{s}^{-1}$,  meaning that background photons coupled from nearby fibers or from the metropolitan environment is minimal.
Overall, the background photon rate introduced by our apparatus is of the same order of the dark count rate of our SPD, and is expected not to significantly affect the QBER.\\ 

\section*{Discussion}
We realised a setup suitable for TF-QKD and characterised it over a \SI{206}{km}-long deployed fiber with \SI{65}{dB} of optical loss. This is the first design for a TF-QKD setup on a commercial network under normal operating conditions and with these losses. This required solution of several fresh challenges for the real-world implementation of TF-QKD, such as a considerably higher attenuation  than on spooled fiber (\SI{0.3}{dB/km} on our setup), autonomous and remote-controlled operation of the equipment at the Alice and Bob terminals,  which was deployed in telecom shelters in a non-controlled environment, and a considerable unbalance (\SI{22}{km}) in the interferometer arms.
Even under these conditions, we ensured the phase coherence of interfering lasers   over hundreds of milliseconds, i.e. 1000 times more than what reported so far in laboratory trials. Furthermore, besides temperature, acoustic and seismic noise on the fibers, our scheme also compensates non-stationary events such as those due to  antropic activities, and is expected to be  robust against further up-scaling of the infrastructure in terms of length, attenuation and phase noise. \\
The key points of our technique lie in the use of ultrastable lasers with several hundreds of kilometers of coherence-length, which is mandatory considering  the constraints set by the network topology  which prevents from realizing perfectly-balanced interferometers, and  the simultaneous transmission of separate signals for the fiber noise detection and the key streaming. On one hand, this enables to keep the QBER to manageable levels thanks to a tighter control of the phase; on the other hand, it allows more advantageous duty cycles for   the quantum communication. In our experiment we were able to maintain $\sigma_\varphi = $  \SI{0.13}{rad}, corresponding to a QBER of 0.5\%, for about \SI{100}{ms}. Both aspects concur to increase the effective key rate, which is a major advantage especially on long haul networks, where  rather low rates of a few \SI{}{kb/s} must be already taken into account due to the fiber losses \cite{wang}.\\
We note that the realised scheme allows rejection of the service fiber noise as well. While previous implementations focus on the strategies for mitigating the noise on the QKD fiber, the issue of noise on the service fiber was only marginally addressed so far \cite{chen,liu}, proposing the Doppler noise cancellation  \cite{williams} as an effective solution. However, we note that this approach is bandwidth-limited by the time needed by the light to travel the fiber, and would  leave several radians of uncompensated phase fluctuations in a realistic case \cite{williams}. 
In our multiplexed scheme, on the contrary, the noise detection is performed upon recombination in Charlie, and the correction can be applied without delay, thus ensuring a higher suppression.
Another approach exploits a Sagnac-interferometer-based configuration \cite{zhong}. We note that these schemes  would suffer from the same limitations as the Doppler stabilization \cite{clivatiSagnac}. On the basis of the results obtained in this work, we also foresee the Rayleigh effect  as a major source of background photons in a Sagnac loop. Rayleigh scattering poses a limitation on the maximum distance achievable when time division multiplexing strategy is employed \cite{chen}, while the wavelength division multiplexing strategy we are proposing is free from this limitation.\\
The proposed scheme can be directly implemented in real quantum communication systems. We underline that, together with phase fluctuation, there are other non-ideal behaviors in the encoding pattern (modulation and phase) of the QKD lasers at the remote terminals that may increase the QBER (i.e. reduction of the visibility interference). Among these, are the relative jitter of the clocks referencing the patterns in Alice and Bob, and the pulses’ arrival time in Charlie which is in principle affected by the varying delay added by the fibers. In this perspective, we note that our scheme supports a common clock signal to be delivered to the terminals trough the service fiber. The additional timing delay introduced by the QKD fibers is \SI{<1}{ps}. Thus, provided that the modulation patterns are initially matched to account for the different lengths of the interferometer arms, the fibers delay is not expected to affect the visibility of the interference even at the high modulation rate of $\sim$\SI{1}{GHz}. However, the multiplexed approach used to stabilize the optical phase could be further exploited to compensate for the overall jitter of the service and QKD fibers. 
Finally, we note that the results of this work are not related to any specific system architecture, and describe useful  strategies for a variety of QKD protocols, significantly contributing to the route towards quantum secure communications in a real field.


\section*{Acknowledgements}
We thank  Consortium Top-IX, especially Matteo Frittelli and Alessandro Galardini, for technical assistance on deployed fibers and helpful discussions on the network design; PPQSense for the fruitful collaboration in customizing the lasers drivers.\\
M.L. has carried out part of this work at Toshiba Europe Ltd.\\
We acknowledge funding from projects Nato G5263; EMPIR-18SIB06-TIFOON, which has received funding from the EMPIR programme co-financed by the EMPIR Participating States and from the European Union's Horizon 2020 research and innovation programme; OpenQKD (Grant Agreement No 857156) and QCall (675662), both of which were also funded by the EU’s Horizon 2020 research and innovation programme. 
\section*{Author contributions}
M.L. and I.P.D. had the initial idea for the work.\\
C. C., S. D., A. Mu., D. C. designed and developed the experimental setup and  performed the measurements in the classical regime.\\
A. Me., S. V., I. P. D.  contributed QKD expertise and performed the measurements in photon-counting regime.\\
F. L. and M. G. provided fundamental knowledge and insight on key metrological and QKD aspects  of the work and substantially contributed to the experimental design.\\
M. L., M. P.,  Z. L. Y., A. J. S.  contributed background in TF-QKD and to the design of the experimental setup. \\
I. P. D., D. C. coordinated the project and supported the experimental activity. \\
All authors discussed the experimental data and contributed to the analysis of results. \\
C. C., with contribution from all authors, wrote the paper.\\
\section*{Methods} 
\subsection*{The optical fiber network}
The fibers used for this experiment are part of the Italian quantum backbone, which provides atomic clock  dissemination services to scientific and commercial users of the Country  \cite{clivatiVLBI}. These services, as well as standard data traffic for the remote control of the equipment at the network nodes, require bidirectional optical transmission. To ensure compatibility with QKD, we migrated the traffic to the single service fiber, using different wavelengths for the two directions of propagation. In particular, time/frequency dissemination services used channel 30 and 31 of the International Telecommunication Union (ITU) grid, corresponding to a wavelength of \SI{1553.33}{nm} and \SI{1552.52}{nm} respectively. Remote control was established over channels 28 and 29 (wavelengths \SI{1554.94}{nm} and \SI{1554.13}{nm}). Standard DWDM multiplexers were used to combine and separate channels 28-31 at the network nodes, while the sensing and reference lasers travelled through the unfiltered ports, which brought $\sim$\SI{2}{dB} additional losses each. 
\subsection*{Reference and sensing lasers}
The reference laser is a fiber laser at \SI{1542.14}{nm} (channel 44 of the \SI{100}{GHz}-DWDM grid), frequency stabilized to an ultrastable Fabry-Perot cavity with a Finesse of 120'000 using the Pound-Drever-Hall technique \cite{clivatiuffc}. The resulting linewidth is \SI{1}{Hz}  and the short-term instability is \SI{2e-15}{}. The cavity is made of ultra-low expansion glass, housed in high vacuum and placed on a platform for passive seismic noise damping. We used an external acousto-optic modulator (AOM) as a fast actuator to lock the fiber laser to the cavity. The achieved bandwidth of \SI{200}{kHz} is limited by the internal delay of the AOM and by its driver. Although diode lasers offer much higher control bandwidths, their phase noise is higher as well, which deteriorates the phase coherence and makes the use of fiber laser preferable for this application.  
The reference laser is a diode laser at \SI{1543.33}{nm}. This wavelength lies in the middle between the channels 42 and 43 of the \SI{100}{GHz}-DWDM grid, and is a standard of the advanced \SI{50}{GHz}-DWDM grid. We virtually phase-locked it to the reference laser using an optical comb as a spectral bridge. The comb is an octave-spanning Er:fiber femtosecond frequency comb with \SI{250}{MHz} repetition rate  whose spectral emission is centered around \SI{1560}{nm}. Following the technique described in \cite{telle}, we detected the beatnote of both lasers with the comb and measured their phase-difference on a mixer. This enabled us to detect the relative phase  between the two, which could not be  directly measured because of the large spectral separation. The phase error was then used to phase-lock the sensing laser acting on its current. 
The offset-lock of  the two beatnotes is preferable than the use of independent optical cavities, because it ensures a tight phase-coherence between the reference and sensing lasers, which mitigates the impact of self-delayed lasers noise on the QKD interference pattern. This is possible because the self-delayed noise of the sensing laser is detected alongside with the optical fiber length variations and equally cancelled by the stabilization loop. As long as this noise is common with the reference laser, it is rejected from  the QKD interference as well. An opposite mechanism would take place using independent lasers, in which case the self-delayed noise of the sensing laser would be written onto the QKD interference. 
A tight phase relation between the two lasers could be maintained  even without using a frequency comb, relying on fast electro-optic modulators and sideband-locking  \cite{santagata}.
\subsection*{Phase lock of the slave diode lasers}
About \SI{20}{\micro W} of the reference laser power was launched in the service fiber towards  Alice and Bob terminals. Here, commercial diode lasers with an optical power \SI{<10}{mW} were phase-locked to it. To this purpose, we detected the beatnote between incoming and local light on a fast photodiode and compared it to a stable radio-frequency oscillator on a mixer in a quadrature condition. The phase error signal was processed by a proportional-integrative-derivative controller acting on the  laser  current. The phase-locked-loop bandwidth of \SI{1}{MHz} results from a combination of the  frequency vs current response of the diode  and the current driver's bandwidth.\\
The local oscillator for the phase-locked loop in Alice (Bardonecchia) was referenced to a  \SI{10}{MHz} signal disseminated with a White-Rabbit Precise Time Protocol over the service fiber, with short-term frequency stability of \SI{1e-11}{} \cite{serrano,dierikx}. In Bob (Santhià), where this service was not activated, we used a Rubidium oscillator with short term stability of \SI{1e-12}{}.
\subsection*{Cancellation of the fiber phase noise}
The launched power of the sensing laser  was about \SI{1}{mW} into each arm of the interferometer. We stabilized the relative phase between the two return beams incoming in Charlie after travelling the path toward the remote terminals and back. To do so, we spilled out a portion of the sensing radiation before sending it to the remote terminals, and  we  detected the beatnote with the return signal on each arm. The resulting beatnotes at \SI{40}{MHz}, the AOMs and AOMa frequencies,  are down-scaled by a factor of 10 and phase-compared on a mixer in quadrature condition. The resulting error signal  drives a proportional-integrative controller which adjusts the phase of both the sensing and QKD lasers by acting on the  frequency of AOMa. 
The bandwidth of the phase-locked loop is limited to \SI{50}{kHz}, which is large enough to fully compensate the acoustic noise introduced by the fiber and the residual self-delayed laser noise. 

\subsection*{The normalised interference pattern}
The  pattern produced by the classical interference of the QKD lasers in Charlie  is  modelled as $I=2 I_0(1+\cos\varphi)$ where  $I_0$ is the lasers' intensity, assumed equal, and $\varphi$  their phase difference. In the experiment, we equalised the intensities of the two beams and aligned their polarization  to maximise the contrast. We then considered the normalised interference pattern $\bar{I}=I/4I_0 = \cos^2( \varphi/2)$, which can be regarded as  the classical counterpart of  a single photon interference. Operating the interferometer in a condition where $\varphi= 0$ or $ \varphi=\pi $ corresponds to the case where all photons  would be routed to one or the other port of the beamsplitter, i.e. operation in a dark port configuration.
On the contrary, when the interferometer operates at $\varphi= \pi/2 $ or $ \varphi=3\pi/2$, the probability of being detected on one or the other port are equal. In this condition, the phase fluctuations are directly mapped into  intensity fluctuations,  as in Fig. \ref{fig:vis}b.  In our experiment, the residual phase fluctuations and its deviation are calculated from $\bar{I}$ inverting the related equation. 

\subsection*{Statistical methods} 
The variance of the phase  $\sigma^2_\varphi$, or its corresponding deviation $\sigma_\varphi$ at a given measurement time $t_\text{a}$ can be directly calculated from time domain data, or as the integral of the power spectrum, which in turns is calculated from instantaneous phase data.
In our experiment, we adopted both methods. First, we  computed the Welch periodogram of the phase $S_\varphi(f)$, as retrieved from the interference pattern, and integrated it between the Fourier frequencies $f=1/t_\text{a}$ and $f=f_\text{s}/2$, where $f_\text{s}$ is the sampling rate:
\begin{equation}
\label{eq:sigma_sp}
   \sigma^2_\varphi =\int_{1/t_\text{a}}^{f_\text{s}/2}S_\varphi(f) \,df
\end{equation}
We note that $f_\text{s}$ must be at least twice as large as the noise bandwidth of the observed pattern to fulfil the Nyquist-Shannon sampling theorem  \cite{shannon}. 
In addition, we evaluated 
the phase variance over the time $t_\text{a}$ by dividing the data set, composed of $N$ phase samples $\varphi_j$ and with total duration $T=N/f_\text{s}$, in subsets of $n$ points, where $n \approx N t_\text{a}/T $. We then computed the standard deviation of each subset and averaged over the number of subsets $i \approx N/n$:
\begin{equation}
\label{eq:sigma_si}
    \sigma^2_\varphi =  \langle \frac{1}{n-1}\Sigma_{j=0}^n (\varphi_j-\bar{\varphi})^2 \rangle_i
\end{equation}
where $\bar{\varphi}$ is the average phase over each subset.
We verified that both methods lead to the same result. \\
The obtained parameter is used to evaluate the QBER. 
When the interferometer is in the $\varphi\approx0$ condition and all the counts are expected to be on a single detector, the  QBER represents the probability of having clicks on the complementary one. The contribution to the QBER   from decoherence  is hence calculated from the phase noise of the system according to the relation  
\begin{equation}
\label{eq:qber}
    e =\int \left( 1-\cos^2 ( \varphi/2) \right) P(\varphi)\, d\varphi =\int \sin^2 ( \varphi/2) P(\varphi)\, d\varphi
    \label{eq:qber}
\end{equation} 
  As long as $\varphi \approx 0$, which is the only interesting case in practice, it can be seen that Eq. \ref{eq:qber} is simplified to $e=\sigma^2_\varphi/4$, where $\sigma^2_\varphi$ is calculated from Eqs. \ref{eq:sigma_sp} or \ref{eq:sigma_si}. 
  In Fig. \ref{fig:sigma} and throughout the text, we use this relation to evaluate the QBER.
\subsection*{Single photon detectors}
We employed a commercial fiber-coupled InGaAs/InP avalanche detector (Id Quantique ID230). The detector mounts a Stirling cooler that enables to cool down to \SI{-90}{\celsius}, reducing the dark counts related to the detection process to a negligible level. It operates in free-running mode, enabling asynchronous photon detection with \SI{150}{ps} timing resolution, in a spectral bandwidth ranging from \SI{900}{nm} to \SI{1700}{nm}. The quantum efficiency is variable up to 25\%  and its dead time can be adjusted from \SI{2}{\micro s} to \SI{100}{\micro s}.

\subsection*{Optical Filtering}
Our technique is based on the  transmission in the same fiber of two separate signals, the QKD lasers and the sensing laser, both in the C-band. Besides the issues related to  nonlinear effects which   generate background photons in the QKD lasers band, a key aspect  is the  efficient separation of the two signals in Charlie, to avoid that photons outside the QKD laser band reach the detector. This is primarily obtained with two cascaded \SI{100}{GHz}-DWDM filters, each featuring \SI{60}{dB} rejection at an offset of \SI{1.5}{nm} from the central wavelength. However, the performances of standard telecom devices drop beyond \SI{1300}{nm}, allowing a non-negligible power from the amplified spontaneous emission of the sensing laser, which extends to a wavelength of \SI{1200}{nm},  to impinge the detectors. This was suppressed by placing a pair of additional free-space filters in front of the detectors, with nominal \SI{50}{dB} rejection over the visible and near-IR band. Their  \SI{10}{nm} bandwidth, combined with the stronger selectivity of DWDM filters, ensured efficient filtering of the quantum photons. The overall losses of cascaded filtering stages  amount to \SI{2}{dB}, which is the result of the 84\% transmissivity of the free-space filters and the coupling losses in the fiber/air interfaces. 

\end{document}